\documentclass[11pt]{article} 
\usepackage{amsmath,amsthm,amscd,amssymb}
\usepackage{latexsym}
\usepackage{epsf}
\usepackage{epsfig}

\newcommand{\fr}{{^F\hspace{-.02in}R}}

\theoremstyle{plain}

\theoremstyle{definition}

\theoremstyle{remark}

\newtheorem{case[theorem]}{Case}

\title{Tidal Dynamics of Relativistic Flows Near Black Holes}
\author{C. Chicone\\Department of Mathematics\\University of
Missouri-Columbia\\Columbia, Missouri 65211, USA
\and B. Mashhoon\thanks{Corresponding author. E-mail:
mashhoonb@missouri.edu (B. Mashhoon).} \\Department of Physics and
Astronomy\\University of Missouri-Columbia\\Columbia, Missouri 65211, USA}
\date{ }
\begin{document}
\maketitle

\begin{abstract}
We point out novel consequences of general relativity involving tidal dynamics of
ultrarelativistic relative motion. Specifically, we use the generalized Jacobi equation
and its extension to study the force-free dynamics of relativistic flows near a massive
rotating source. We show that along the rotation axis of the gravitational source,
relativistic tidal effects strongly \emph{decelerate} an initially ultrarelativistic flow 
with respect to the ambient medium, contrary to Newtonian expectations.
Moreover, an initially ultrarelativistic flow perpendicular to the axis
of rotation is strongly \emph{accelerated} by the relativistic
tidal forces. The astrophysical implications of these results for jets and
ultrahigh energy cosmic rays are briefly mentioned.
\end{abstract}

\begin{description}\item[PACS numbers:] 
04.20.Cv, 97.60.Lf, 98.58.Fd, 98.70.Sa
\item[Keywords:] ultrarelativistic particles, tidal acceleration, 
tidal deceleration
\end{description}

\section{Introduction}\label{s1}
This paper is about the motion of free test particles in the background
gravitational field of a Kerr black hole. The geodesics of Kerr spacetime are well
known and have been extensively studied in standard systems of coordinates such as
the Boyer-Lindquist coordinate system. The Boyer-Lindquist coordinates are
adapted to the stationary and axisymmetric nature of the source and are Minkowskian
(expressed in spherical polar coordinates) infinitely far from the black hole.
We
are interested however in the \emph{relative} 
motion of free nearby particles; as explained
in detail in~\cite{1}, from the standpoint of relativity theory the study of relative
motion is certainly preferred, since it is in keeping with the spirit of the theory.
Moreover, relative motion has direct observational significance. Thus theoretical
results obtained in the study of relative motion can be directly compared with
suitable experimental data~\cite{1,2}.

To study the relative motion of free particles, we need a reference particle
whose motion is well known. For the purposes of the present work, we select the
free motion of particles on radial escape trajectories along the rotation axis
of the Kerr black hole. Thus at a given initial position along a radial
direction in Schwarzschild spacetime or the rotation axis in Kerr spacetime,
the reference particle has an escape velocity such that the particle eventually
reaches infinity with zero kinetic energy. We then study the relative motion of
the relativistic particles that start from the same location as the reference
particle and travel in different directions. The origin of these relativistic
particles on the axis of rotation will not be discussed in this paper. We will
simply assume that such particles exist near the poles of the Kerr black hole.
The nature of the central engine and the details of the accretion phenomena
that could generate such relativistic particles are beyond the scope of this
work. Instead, we concentrate on the gravitational dynamics of these particles
relative to the reference particle. Moreover, it turns out that the existence of event 
horizons has no bearing on our main results; therefore, our treatment 
applies equally well to test particles in the exterior of a gravitational 
source whose field is adequately described by the exterior Kerr spacetime. 
However, our results are in general observationally significant only in the 
case of highly collapsed gravitational sources such as neutron stars and 
black holes.

To characterize the relative motion, we establish a quasi-inertial 
coordinate
system along the worldline of the reference particle. This Fermi normal
coordinate system~\cite{3} is valid in a cylindrical region of radius $\mathcal{R}$ 
along the
worldline of the reference particle. Here $\mathcal{R}$ is 
the radius of curvature
of the spacetime manifold. The geodesic equations of motion for the free
particles are then integrated in this Fermi coordinate system subject to the
limitation that the relative distance must remain well within the admissible
range of Fermi coordinates. 
The results of such an investigation of the relative motion of free particles 
would
naturally depend on the choice of the reference particle. For instance, a
separate investigation would be necessary if the reference trajectory is an
orbit in the equatorial plane of the Kerr black hole. In fact, a
judicious choice of the reference particle would be essential to explain a
particular observational result. Moreover the relative distance in our approach
should be less than $\mathcal{R}$.

We find that ultrarelativistic flows in the gravitational field of collapsed  configurations can
exhibit purely non-Newtonian behavior due to  general relativistic tidal 
effects. These
include the phenomenon of the tidal  deceleration
along the axis of rotation and the phenomenon of tidal acceleration
perpendicular to the axis of rotation, which follow from the integration
of the generalized Jacobi equation \cite{1,2}. We study this equation and its
extension in this paper and compare our results with the corresponding nonrelativistic
tidal effects. In the  following we use units such that $c=1$.
The signature of the spacetime metric is assumed to be $+2$.

Imagine a geodesic worldline in the gravitational field of an 
external source. Let $u^\mu=dx^\mu /d\tau$ be the four-velocity 
vector of this {\it reference} geodesic and $\tau$ be its proper 
time. A neighboring geodesic with an arbitrary velocity relative to 
the reference geodesic is connected to it by the deviation vector 
$\xi ^\mu (\tau )$ such that $\xi^\mu u_\mu=0$. The generalized 
Jacobi equation \cite{1} is given by the geodesic deviation equation to first 
order in $\xi^\mu$, namely,
\begin{eqnarray}\label{eq1}
\nonumber &&\frac{D^2\xi^\mu}{D\tau^2}+R^\mu _{\;\; \rho \nu \sigma} u^\rho 
\xi^\nu u^\sigma \\
\nonumber &&{}
+\Big(u^\mu +\frac{D\xi^\mu}{D\tau }\Big) 
\Big( 
2R_{\zeta \rho \nu \sigma}u^\zeta \frac{D\xi^\rho}{D\tau }\xi ^\nu 
u^\sigma +\frac{2}{3} R_{\zeta \rho \nu \sigma}u^\zeta \frac{D\xi 
^\rho}{D\tau}\xi^\nu \frac{D\xi^\sigma}{D\tau}
\Big)\\
&&{}+2R^\mu _{\;\; \rho \nu \sigma} \frac{D\xi^\rho}{D\tau 
}\xi^\nu  u^\sigma +\frac{2}{3}R^\mu_{\;\; \rho \nu 
\sigma}\frac{D\xi^\rho}{D\tau}\xi^\nu \frac{D\xi^\sigma}{D\tau}=0.
\end{eqnarray}
Here $D\xi^\mu /D\tau =\xi^\mu_{\;\; 
;\nu}u^\nu$ is the covariant derivative of $\xi^\mu$ along the 
reference worldline. Let $\lambda^\mu _{\;\;(\alpha)}$ be an 
orthonormal tetrad frame that is parallel propagated along the 
reference trajectory such that $\lambda^\mu_{\;\;(0)}=u^\mu$. Then, 
$\xi^\mu =X^i\lambda^\mu_{\;\; (i)}$ and equation \eqref{eq1} may be 
written in terms of Fermi coordinates $(T,{\bf X})$, where $T=\tau$ 
along the reference geodesic, as \cite{1,2}
\begin{eqnarray}\label{eq2}
\nonumber &&\frac{d^2X^i}{dT^2}+\fr_{0i0j}X^j+2\,\fr_{ikj0}V^kX^j\\
&& 
+(2\,\fr_{0kj0}V^iV^k+\frac{2}{3}\,\fr_{ikj\ell}V^kV^\ell 
+\frac{2}{3}\,\fr_{0kj\ell}V^iV^kV^\ell )X^j=0.
\end{eqnarray}
Here $V^i=dX^i/dT$, $|\mathbf{V}|< 1$ along the reference geodesic and
\begin{equation}\label{eq3}
\fr_{\alpha \beta \gamma \delta 
}(T)=R_{\mu \nu \rho \sigma }\lambda^\mu_{\;\; (\alpha )}\lambda^\nu 
_{\;\;(\beta)} \lambda^\rho _{\;\; (\gamma )}\lambda^\sigma 
_{\;\;(\delta )}
\end{equation}
is the projection of the Riemann 
curvature tensor on the orthonormal tetrad of the reference observer along its 
worldline. If the relative velocity is much smaller than the speed of 
light, $|\mathbf{ V}|\ll 1$, the velocity-dependent terms in \eqref{eq1} 
and \eqref{eq2} may be neglected and the generalized Jacobi equation 
reduces to the standard Jacobi equation that expresses the linear 
evolution of the deviation between two neighboring geodesics that 
have negligible relative motion. This relative motion generally 
evolves slowly on the scale of the radius of curvature $\mathcal{R}$, which 
is defined such that $\mathcal{R}^{-2}$ is the supremum of 
$|\fr_{\alpha \beta \gamma \delta} |$; therefore, $|{\bf X}|\ll 
\mathcal{R}$ in \eqref{eq2},  since higher-order terms in the deviation 
equation have been neglected. In the generalized Jacobi equation, the 
relative velocity could approach the speed of light; therefore, the 
relative motion could evolve rapidly thus further restricting the temporal 
domain of validity of equation \eqref{eq2}.

Equation~\eqref{eq2} can be derived in a straightforward manner
using the reduced geodesic equation in Fermi coordinates. That is,
starting from the geodesic equation for a free particle in 
arbitrary coordinates $\hat x^\mu=(\hat t, \hat x^i)$,
\begin{equation}\label{mu:4}
\frac{d^2\hat{x}^\mu}{ds^2} +\hat{\Gamma}^\mu_{\alpha\beta}
 \frac{d\hat{x}^\alpha}{ds}\frac{d\hat{x}^\beta}{ds}=0,
\end{equation}
where $ds^2=\hat{g}_{\mu\nu}d\hat x^\mu d\hat x^\nu$, one
can simply derive the reduced geodesic equation~\cite{1} 
\begin{equation}
\frac{d^2\hat{x}^i}{d\hat{t}^2}
-(\hat\Gamma^0_{\alpha\beta}\frac{d\hat{x}^\alpha}{d\hat{t}}\frac{d\hat{x}^\beta}{d\hat{t}})\frac{d\hat{x}^i}{d\hat{t}}
+\hat\Gamma^i_{\alpha\beta}\frac{d\hat{x}^\alpha}{d\hat{t}}\frac{d\hat{x}^\beta}{d\hat{t}}=0.
\end{equation}
Specializing this equation now to the Fermi coordinates established
along the reference trajectory, i.e. $(\hat t, \hat x^i)\mapsto (T,X^i)$,
and using 
\begin{eqnarray}
\label{metrict} 
\nonumber g_{00}&=&-1-\fr_{0i0j}(T)\,X^iX^j+\cdots,\\
\nonumber g_{0i}&=&-\frac{2}{3}\,\fr_{0jik}(T)\, X^jX^k+\cdots,\\
g_{ij}&=&\delta_{ij}-\frac{1}{3}\,\fr_{ikj\ell}(T)\, X^kX^\ell+\cdots,
\end{eqnarray}
one arrives at the tidal equation for the \emph{relative} motion,
since in these coordinates the reference particle remains fixed at 
$\mathbf{X}=0$. Restricting our attention to the terms given explicitly
in~\eqref{metrict} and thus neglecting higher-order terms of the metric
in the spatial Fermi coordinates, the tidal equation reduces to the generalized
Jacobi equation~\eqref{eq2}. We study the main physical consequences
of this equation in the field of a Kerr black hole 
(see also~\ref{appen:A}). 
Higher-order tidal accelerations are discussed in~\ref{appen:B}. 

Tidal deceleration along the rotation axis is discussed in the next section. 
The non-Newtonian character of our results is emphasized in section~\ref{s3}.  
The analysis of the generalized
Jacobi equation is extended to the three-dimensional case in
section~\ref{s5}. Tidal acceleration normal to the rotation
axis is discussed in section~\ref{s6}.
Finally, section~\ref{s7} contains a brief 
discussion of our results.

\section{Tidal deceleration}\label{s2}
Imagine a Kerr source of mass $M$ and angular momentum 
$\mathbf{J}=Ma\, \hat{\mathbf{z}}$ at the 
origin of an asymptotically inertial coordinate system $(t,x,y,z)$. 
Here $a>0$ is the specific angular momentum of the source. 
The metric of Kerr spacetime can be written in Boyer-Lindquist coordinates
as 
\begin{eqnarray}\label{eq:neq4}
\nonumber ds^2&=&-dt^2+\Sigma (\frac{1}{\Delta}dr^2+d\vartheta^2)+(r^2+a^2)\sin^2\vartheta\,d\phi^2\\
&&{}+2GM\frac{r}{\Sigma}(dt-a\sin^2\vartheta\,d\phi)^2,
\end{eqnarray}
where $\Sigma=r^2+a^2\cos^2\vartheta$ and $\Delta=r^2-2GMr+a^2$.
Asymptotically ($r\to \infty$), the spherical polar coordinates $(r,\vartheta,\phi)$
are related to the Cartesian coordinates $(x,y,z)$ in the standard manner. 
In this paper, we are interested in the motion of free
test particles in the stationary and axisymmetric exterior
Kerr spacetime. Let us first consider geodesic motion
along the axis of rotation of the source,  i.e. the $z$-axis. In terms of the
Boyer-Lindquist temporal coordinate $t$ and radial coordinate 
$r$, the motion of a free test particle along the rotation axis is 
given by
\begin{equation}\label{eq4}
\frac{dt}{d\tau} =\gamma \frac{r^2+a^2}{r^2-2GMr+a^2},\quad 
\frac{dr}{d\tau }=\pm \sqrt{\gamma^2-1+\frac{2GMr}{r^2+a^2}},
\end{equation}
where $\tau $ is 
the proper time along the path and $\gamma >0$ is a constant of 
integration. For $\gamma <1$, the test particle cannot escape the 
gravitational attraction of the source. When the test particle can 
escape to infinity, $\gamma \geq 1$ and $\gamma$ is its Lorentz 
factor at spatial infinity. In the limiting case of $\gamma =1$, the 
test particle approaches infinity with zero speed. We assume in the 
following that there are many such outward moving test particles with $\gamma=1$ in 
the neighborhood of the rotation axis of the Kerr source forming an 
ambient medium around the rotating source. Relative to this ambient 
medium, we wish to study the motion of an ultrarelativistic jet 
emitted outward from a region near the source along the axis of 
rotation.

Consider a jet clump starting from $r_0$ and moving rapidly along the $z$-axis
according to equation \eqref{eq4} with $\gamma >1$. This relative motion is properly
described in the first approximation using the generalized Jacobi equation \eqref{eq1}
expressed in Boyer-Lindquist coordinates. To compare the consequences of such an
equation with astronomical data regarding the motion of jets, it is appropriate to
transform this equation to a quasi-inertial Fermi normal coordinate system that is
established along the reference trajectory. The {\it explicit} relationship between the
Boyer-Lindquist coordinate system of Kerr spacetime and the Fermi system is not needed
for the derivation of the equation of relative motion in Fermi coordinates \eqref{eq2}.
We therefore proceed keeping in mind that the radial coordinate $r$ of the jet clump is
a function of the Fermi coordinates, $r=r (T,Z)$, but that this relationship shall
remain implicit throughout this work.

Along the worldline of a generic outward moving test particle with $\gamma =1$ on 
the rotation axis, we establish a Fermi coordinate system $(T, X, 
Y,Z)$ based on an orthonormal nonrotating (i.e. Fermi-Walker 
transported) tetrad frame. The free test particle follows a geodesic; 
therefore, Fermi-Walker transport reduces
to parallel transport in this case. 
The Fermi coordinates of the reference particle
are then $(\tau, 0, 0, 0 )$, 
i.e. the reference particle is at the spatial origin of
the Fermi coordinates.
The quasi-inertial Fermi system is the one 
appropriate for the comparison of the theory with observation as well 
as Newtonian physics \cite{3}. In fact, the speed of a jet is 
determined by monitoring the motion of a jet clump with respect to 
neighboring ``fixed" features of the ambient medium \cite{4}. The 
motion of the jet clump relative to the reference particle in the Fermi 
system is given by the tidal equation \cite{1}

\begin{equation}\label{eq5}
\frac{d^2Z}{dT^2}+k(1-2V^2)Z+O(Z^2)=0,
\end{equation}
where $V=dZ/dT$ and the curvature $k=\fr_{0303}$ is 
given by
\begin{equation}\label{eq6}
k=-2GM\frac{r(r^2-3a^2)}{(r^2+a^2)^3}.
\end{equation}
Here $r$ is the Boyer-Lindquist radial coordinate of the \emph{reference}
particle and $r(T)$ can be determined from the integration of the
second equation in~\eqref{eq4} with $\gamma=1$, $\tau=T$ and $dr/dT>0$.
We will be 
concerned with regions along the rotation axis such that $r^2\gg 
3M^2\geq 3a^2$, so that $k<0$. Solving equation \eqref{eq4} with 
$\gamma =1$ and $\tau\to T$ for $r(T)$ and substituting this in 
equation \eqref{eq6} gives us the appropriate $k(T)$ to use in the 
tidal equation \eqref{eq5}. Let us note that equations~\eqref{eq4}--\eqref{eq6} 
are valid for any free reference particle
moving along the rotation axis of the Kerr source.

The specific angular momentum of the source $J/M=a$ appears in 
equations~\eqref{eq4} and~\eqref{eq6} as $a^2$, so that
these equations are invariant under the transformation $a\mapsto -a$.
This implies that the sense of rotation of the black hole is of no
consequence for the phenomena described in this section. This
is consistent with the double-jet structure observed
in astrophysics. It is therefore sufficient to treat only the
motion of particles moving outward above the north pole of the black hole.

It is important to mention here that the main results of this work are independent of
the precise nature of the massive source. For the sake of concreteness, we have assumed
that the source is a Kerr black hole with $a\leq GM$. Moreover, we will concentrate on
a jet clump launched from $r_0\gg GM$ and moving along the positive $z$-axis. However,
equations \eqref{eq4}--\eqref{eq6} depend upon $a$ in terms of $a^2/r^2$, which is very
small for $r\ge r_0\gg a$. In fact, our main results are insensitive to the ratio $a/(GM)$.
That is, our main conclusions remain the same even if the source is spherically
symmetric $(a=0)$ and the jet clump is launched along {\it a radial direction}.

The tidal equation~\eqref{eq5} in principle contains an infinite 
series in powers of the quantity $Z/\mathcal{R}$, where $\mathcal{R}$ 
is the effective radius of curvature such that 
$\mathcal{R}^{-2}=|k|$. For regions of interest in this work 
$k\approx -2GM/r^3$ and hence
\begin{equation}\label{eq7}
\mathcal{R}\approx \sqrt{\frac{r^3}{2GM}}.
\end{equation}
Starting from $r=r_0$ corresponding to $Z=0$ at $T=0$, we integrate equation
\eqref{eq5} with a given initial $V_0=V(0)>0$. 
For a highly relativistic jet clump, the
initial Lorentz factor $\Gamma_0=(1-V_0^2)^{-\frac{1}{2}}$ is such that $\Gamma _0\gg
1$. The integration time $T$ must be limited such that $Z(T)\approx T<\mathcal{R}_0$,
where $\mathcal{R}_0$ is given by equation \eqref{eq7} at $r=r_0$. It is important to
note that the endpoint of the integration is only known implicitly in terms of the
Boyer-Lindquist radial coordinate $r$. There is, of course, a definite relationship
between the  Boyer-Lindquist coordinates and the corresponding Fermi coordinates. 
However, this relationship remains implicit here; making it explicit would immediately
imply that we know the  tidal equation~\eqref{eq5} to all orders~\cite{1}.
\begin{figure}
\centerline{\psfig{file=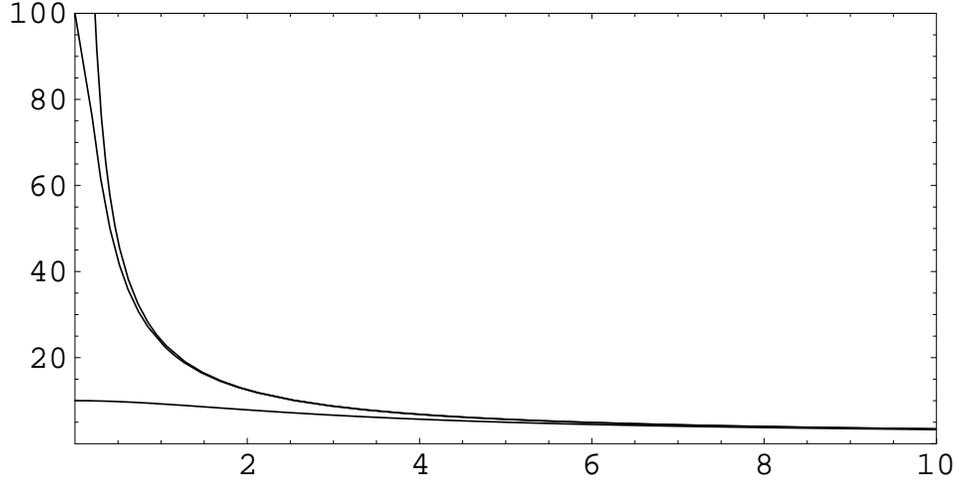, width=30pc}}
\caption{Plot of the Lorentz factor $\Gamma$ versus $T/(GM)$  
based on the integration
of equation~\eqref{eq8} with initial data
$r_0/(GM)=10$, $Z(0)=0$ and $V(0)=\sqrt{\Gamma_0^2-1}\,/\,\Gamma_0$ for 
$a/(GM)=1$. The graph illustrates the deceleration of clumps with
initial Lorentz factors $\Gamma_0=1000$, 100 and 10. 
\label{fig:1}}
\end{figure}
\begin{figure}
\centerline{\psfig{file=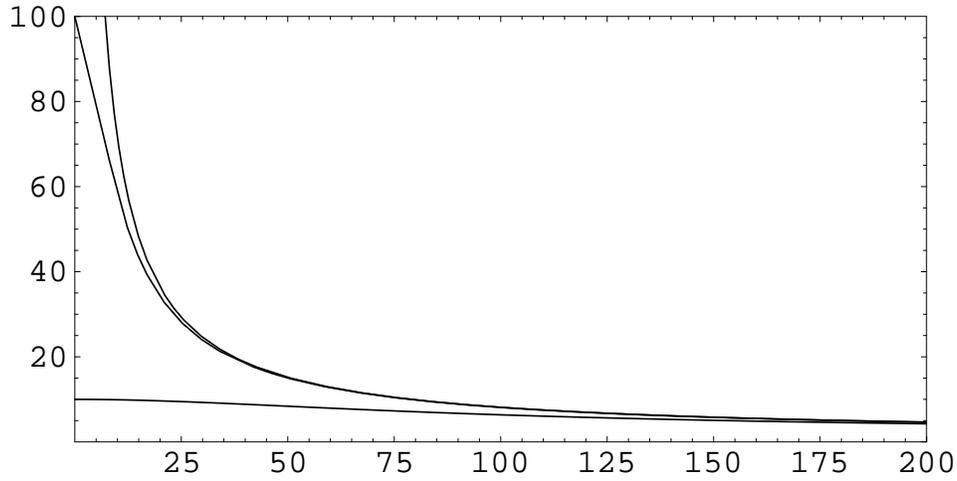, width=30pc}}
\caption{Plot of the Lorentz factor $\Gamma$ versus $T/(GM)$  
based on the integration
of equation~\eqref{eq8} with initial data
$r_0/(GM)=100$, $Z(0)=0$ and $V(0)=\sqrt{\Gamma_0^2-1}\,/\,\Gamma_0$ for 
$a/(GM)=0$. The graph illustrates the deceleration of clumps with
initial Lorentz factors $\Gamma_0=1000$, 100 and 10. 
\label{fig:1a}}
\end{figure}

The first-order approximation 
of the tidal equation corresponds to the generalized Jacobi 
equation,
\begin{equation}\label{eq8}
\frac{d^2Z}{dT^2}+k(1-2V^2)Z=0.
\end{equation}
This equation has solutions corresponding to $V=\pm 
1/\sqrt{2}$, or $\Gamma :=(1-V^2)^{-\frac{1}{2}}=\sqrt{2}$, where the 
jet moves with uniform speed relative to the reference particle. 
For our reference particle with 
$\gamma =1$, these special solutions are attractors \cite{1,2}. 
Starting near the Kerr black hole and moving along the positive 
$z$-axis, an initially
ultrarelativistic jet clump decelerates rapidly relative to the 
reference particle ($\gamma = 1$ ) according to equation~\eqref{eq8}. 
This initial deceleration and approach to
$1/\sqrt{2}\,$ is depicted within the domain of validity of equation \eqref{eq8} in
figures~\ref{fig:1} and \ref{fig:1a} for three clumps  with initial Lorentz factors of 
$\Gamma_0 = 1000$, 100 and 10.

The critical speeds $\pm 1/\sqrt{2}$ that appear in equation~\eqref{eq8}
correspond to the critical Lorentz factor $\Gamma_c=\sqrt{2}$. 
A particle of mass $m$ with a Lorentz factor of $\Gamma_c$ has kinetic momentum
$m$, hence its de Broglie wavelength equals its Compton wavelength.
For our present purposes, the motion of a particle is \emph{ultrarelativistic} 
if its corresponding Lorentz factor $\Gamma$ exceeds $\Gamma_c$.
For $\Gamma\ll \Gamma_c$,
the tidal equation has already been extensively studied in 
the field of a Kerr black hole~\cite{5,6}. 

The novel deceleration effect is thus a consequence of 
equation~\eqref{eq2} restricted 
to ultrarelativistic motion along a certain straight line. 
It is therefore interesting to
consider the  application of the generalized Jacobi equation to the motion of 
astrophysical jets \cite{1,2}. Thus imagine a rotating gravitational source with 
a double jet system moving outward along its axis of rotation. We have 
assumed that the Fermi coordinates established along an ambient 
trajectory that lies on the axis of rotational symmetry of the source 
are such that the axis of rotation is the $Z$-axis. Concentrating on 
an ultrarelativistic jet moving along the positive $Z$-axis, one finds that equation 
\eqref{eq2} in this case involves a remarkably strong initial deceleration
followed by a gradual decline toward a terminal speed given by 
$1/\sqrt{2}$. Starting with an ultrarelativistic jet, this terminal 
speed is approached asymptotically after an infinite period of time. 
On the other hand, the generalized Jacobi equation is only valid for a 
rather limited timescale that nevertheless covers the initial rapid drop in velocity.

Figures~\ref{fig:1} and~\ref{fig:1a} illustrate the main result of this paper regarding tidal deceleration;
therefore, it is worthwhile to develop an approximate analytic understanding of how
this tidal deceleration comes about. To fix our ideas, we focus attention in what
follows  on a jet clump  with an initial Lorentz factor
$\Gamma_0\gg 1$ moving along  the positive $z$-direction relative to a free test
particle of the  ambient medium. At $T=0$, $Z=0$, i.e. the jet starts from the 
position of the reference particle on the $z$-axis, where
\begin{equation}\label{eq9}
r\approx \Big( 
r_0^{3/2}+\frac{3}{2}\sqrt{2GM}\, T\Big)^{2/3}
\end{equation}
according to equation \eqref{eq4} with $\gamma =1$ and 
$r(T=0)=r_0\gg 2GM$. It follows that in equation \eqref{eq5},
\begin{equation}\label{eq10} 
k\approx 
-2\frac{GM}{\Big(r_0^{3/2}+\frac{3}{2}\sqrt{2GM}\,T\Big)^2}.
\end{equation}

Starting from equation~\eqref{eq8} and writing $d^2Z/dT^2=VdV/dZ$, it is simple to 
show that equation~\eqref{eq8} can be completely integrated for a constant
curvature $k$; in fact, the phase diagram for this case has been given
schematically in figure~2 of~\cite{1}. 
For $k=k(T)$, however, let us note that 
\begin{equation}\label{mu15}
\frac{d}{dT}\ln \Big(\frac{1-2 V^2}{1-2 V_0^2}\Big)
  =4 k(T) V Z.
\end{equation}
Since $k<0$, it follows form equation~\eqref{mu15} that if
$V_0^2<1/2$, $V^2$ increases toward $1/2$, while if  $V_0^2>1/2$, 
$V^2$ decreases toward $1/2$. Integrating equation~\eqref{mu15}, we find
\begin{equation}\label{eq11}
\ln \left( 
\frac{1-2V^2}{1-2V^{2}_0}\right) =4\int^T_0k(T')V(T')Z(T')dT'.
\end{equation}
To find an approximate solution of this equation 
we write equation \eqref{eq11} in the form
\begin{equation}\label{eq12}
\ln 
\left( \frac{1-2\Gamma^{-2}}{1-2\Gamma^{-2}_0}\right) 
=4\int^{T}_0k(T')V(T')Z(T ')dT'.
\end{equation}
Assuming that $\Gamma\gg 1$ and $\Gamma_0\gg 1$,
the left side of equation~\eqref{eq12}
can be approximately written as $2(\Gamma_0^{-2}-\Gamma^{-2})$.
Next, in the 
integrand of equation \eqref{eq12} we let $V(T')\approx 1$ and $Z(T')\approx T'$,
 and note that
\begin{equation}\label{eq13}
\int^{T}_0k(T')T'dT'\approx -\frac{4}{9} \left[ \ln 
(1+\lambda ) +\frac{1}{1+\lambda}-1\right]
\end{equation}
using 
equation \eqref{eq10}. Here $\lambda$ is defined by
\begin{equation}\label{eq14}
\lambda 
=\frac{3}{2}\frac{\sqrt{2GM}\,T}{r_0^{3/2}},
\end{equation}
so that $2\lambda/3=T/\mathcal{R}_0$, using equation~\eqref{eq7}.
For $\lambda \ll 1$, one can show that
\begin{equation}\label{eq15}
\ln (1+\lambda 
)+\frac{1}{1+\lambda }\approx 1+\frac{1}{2}\lambda^2+O(\lambda^3).
\end{equation}
Equations~\eqref{eq12}--\eqref{eq15} thus imply that 
\begin{equation}\label{neweq17}
\Gamma^{-2}\approx \Gamma_0^{-2}+\Big(\frac{T}{\mathcal{R}_0}\Big)^2
\end{equation}
or $V^2\approx V_0^2-(T/\mathcal{R}_0)^2$.
This result provides an approximate solution of equation~\eqref{eq8}
for $\Gamma_0\gg 1$,  $\Gamma\gg 1$ and $T\ll 2\mathcal{R}_0/3$. Indeed, a graph of 
equation~\eqref{neweq17} is almost indistinguishable from the corresponding
numerical solution of equation~\eqref{eq8} presented in figures~\ref{fig:1}
and~\ref{fig:1a}.

To get an estimate of the time $T$ that it 
would take for a highly relativistic jet clump with $\Gamma _0\gg 1$ to 
decelerate to
$\Gamma$, where $\Gamma_0^2\gg \Gamma^2 \gg 1$,
we note that $\lambda \approx 3 /(2\Gamma )$
once we neglect $\Gamma_0^{-2}$ compared with $\Gamma^{-2}$.
 Thus equation~\eqref{neweq17} can be written as
\begin{equation}\label{eq16}
\frac{T}{GM}\approx \frac{1}{\Gamma} \sqrt{\frac{1}{2}\left(
\frac{r_0}{GM}\right)^3}.
\end{equation}
It follows that the deceleration time is essentially
independent of $\Gamma_0$. This remarkable
result is already evident from the form of equation \eqref{eq12}: for $\Gamma_0\gg
\Gamma\gg 1$, the left side of equation \eqref{eq12} is nearly independent of
$\Gamma_0$. In figure~\ref{fig:1}, for instance, $r_0=10\; GM$ and the time it takes for
$\Gamma$ to decrease to $\Gamma \approx 5$ from any initial $\Gamma_0\gg 5$ should be
$\approx 2\sqrt{5}\; GM$ based on equation \eqref{eq16}, in good agreement with the
numerical results of figure~\ref{fig:1}. Thus no matter how relativistic the clump may
be initially at $r_0=10\; GM$, it decelerates to $\Gamma \approx 5$ in a short time of
order $GM$. Moreover, this time is within the limit of validity of the generalized
Jacobi equation, since equation \eqref{eq16} may be written as
\begin{equation}\label{eq17}
\frac{T}{\mathcal{R}_0}\approx \frac{1}{\Gamma} \ll 1.
\end{equation}
The difference between figure~\ref{fig:1} and figure~\ref{fig:1a} is a reflection of the
fact that the curvature $k$ decreases as the inverse cube of the radial distance
$(r\gg GM)$ from the black hole and, using equation \eqref{eq16} for
a fixed $\Gamma$, 
$T^2\propto r^3_0$,
which is reminiscent of Kepler's third law.

Thus far we have
considered only one reference particle of the  ambient medium with $\gamma = 1$. 
Let us now imagine $N$ such particles all with $\gamma = 1$
distributed along the rotation axis of the source and constituting ``fixed"
features of the ambient medium that can be used for reference purposes. 
More
specifically, let the Boyer-Lindquist radial coordinates 
$r_i$, $i = 1,2,\ldots, N$,
of the particles be such that $r_1<r_2<\ldots<r_i<r_{i+1}<\ldots<r_N$; 
moreover, each
interval is small enough that the generalized Jacobi equation is valid within
this interval to a reasonably good approximation. That is, $N$ different Fermi
frames can be established that are centered on the $N$ particles and the
generalized Jacobi equation can be applied separately within each interval 
$[r_i, r_{i+1} ]$, so that if the initial speed of the jet clump relative to 
$r_i$ is
more than the terminal speed, the clump continues to decelerate within this
interval, and so on. 
This is a direct consequence of the remarkable feature of
equations~\eqref{eq4}--\eqref{eq6} noted above: 
they do not explicitly depend on any specific
reference particle with $\gamma = 1$. In this sense, 
the clump decelerates \emph{with respect to the ambient medium} 
toward the terminal speed. As the jet clump approaches the terminal speed, the
first-order tidal term proportional to $1-2V^2$ in equation \eqref{eq5} approaches zero
and hence the higher-order tidal terms, neglected in equation \eqref{eq5}, become
dominant. A complete knowledge of these terms is then required to predict the
subsequent evolution of the flow. It may be that at such a time the clump is
sufficiently far from the black hole that tidal effects become unimportant and plasma
forces take over the dynamics of the clump \cite{2}.

Consider now the tidal equation \eqref{eq5} in the nonrelativistic 
limit $|V|\ll 1$.
The first-order approximation leads in this case to the standard Jacobi equation, i.e.
equation \eqref{eq8} but with $1-2V^2$ replaced by unity. 
As explained before, for a
clump along the rotation axis at $r$ with $r^2 \gg 3a^2$, $k\approx -2GM/r^3$ just as in
Newtonian mechanics. The tidal force of the central source in this case simply tends to
pull the clump and the reference particle apart, leading to a relative tidal {\it
acceleration}. The result of the integration of the standard Jacobi equation is in this
case essentially the same as in the Newtonian theory \cite{5,6}, which is studied in
the next section.

\section{Comparison with Newtonian gravitation}\label{s3}
It is 
useful to compare the main results of the relativistic approach with 
the Newtonian theory of tides, since our main conclusions regarding 
tidal deceleration and terminal speed have no counterparts in the 
Newtonian approach and hence violate our ``nonrelativistic" 
intuition.

Let us consider in the background inertial coordinate 
system $(t,x,y,z)$, the Newtonian equations of motion of the jet 
clump at $r$ and the reference particle at $r_p$ along the rotation 
axis of the source
\begin{equation}\label{eq18}
\frac{d^2r}{dt^2}=-\frac{GM}{r^2},\quad 
\frac{d^2r_p}{dt^2}=-\frac{GM}{r^2_p}.
\end{equation}
We are 
interested in the relative motion; therefore, with $\zeta =r-r_p$,
\begin{equation}\label{eq19} 
\frac{d^2\zeta}{dt^2}=-\frac{GM}{(r_p+\zeta)^2} + \frac{GM}{r_p^2}.
\end{equation}
For $\zeta<r_p$, this equation can be written as
\begin{equation}\label{eq19a} 
\frac{d^2\zeta}{dt^2}=\frac{2GM}{r_p^3} \zeta +O(\zeta ^2),
\end{equation}
which should be compared and contrasted with equation 
\eqref{eq5}. The higher-order Newtonian tidal terms 
are considered in~\ref{appen:B}.

Let us assume as before that the reference particle has 
zero energy, so that its kinetic and potential energies add up to 
zero; then, $(dr_p/dt)^2=2GM/r_p$ can be simply integrated and the 
result is
\begin{equation}\label{eq20}
r_p=\Big( 
r_0^{3/2}+\frac{3}{2}\sqrt{2GM}\, t\Big)^{2/3},
\end{equation}
where $r_p=r_0$ at $t=0$. 
Substituting this result in equation \eqref{eq19}, we 
find to linear order in $\zeta$
\begin{equation}\label{eq21}
\frac{d^2\zeta} {dt^2}-\frac{2GM}{\big( 
r_0^{3/2}+\frac{3}{2}\sqrt{2GM}\,t\big)^2}\zeta =0,
\end{equation}
which can be solved with boundary conditions that $\zeta 
=0$ and $d\zeta /dt=v_0$ at $t=0$. Here $v_0>0$ is the initial speed of 
the jet clump relative to the reference particle. The radial 
Newtonian tidal force tends to pull the particles apart; therefore, 
$\dot{\zeta}=d\zeta /dt$ monotonically increases 
for the duration of the validity of the approximation used here. This 
is the complete opposite of what happens in the relativistic theory
for an initially ultrarelativistic relative speed.

An interesting feature of equation~\eqref{eq8} is therefore that the nature
of the tidal force changes depending on whether $V^2$ is below or above
the critical value $V_c^2=1/2$. 
In fact, for $V^2<1/2$ the tides behave essentially as in
Newtonian mechanics, i.e. the particle accelerates---its speed approaching
$1/\sqrt{2}$ very slowly---relative to the reference particle.
The character of
the tides in a model star falling into a Kerr black hole has been 
investigated in detail in~\cite{5,6} using the Jacobi equation;
the results are similar as in Newtonian gravitation except possibly when
the model star is very close to the horizon of the black hole.
On the other hand,  for $V^2>1/2$ the tides behave in the opposite
way (see figures~\ref{fig:1} and~\ref{fig:1a}). 
 
Comparing the Newtonian behavior with the consequences of equations 
\eqref{eq8}--\eqref{eq10}, we recognize the significance of the 
relativistic tidal acceleration term $2kV^2Z$ for the difference 
between the results of the two theories. Thus tidal deceleration and 
terminal speed exhibited by \eqref{eq8} are due to this 
gravitomagnetic (i.e. motional) tidal effect that is purely relativistic and thus goes
beyond  Newtonian gravitation.
The higher-order tidal terms~\cite{5,6,7}
in equation~\eqref{eq5} are treated using a certain post-Newtonian
approach in~\ref{appen:B}.

\section{Tidal dynamics}\label{s5}
It is interesting to explore the dynamics of the three-dimensional generalized Jacobi
equation \eqref{eq2} using the same reference trajectory \eqref{eq4} as before. The
components of the curvature tensor in the Fermi coordinate system established along
this worldline are given by~\cite{n8}
\begin{eqnarray}\label{eq25}
^F R_{0101}&=&^FR_{0202} =-\frac{1}{2} k,\quad ^FR_{0303} =k,\\
\label{eq26} ^FR_{2323} & =&^FR_{3131}=\frac{1}{2}k,\quad ^FR_{1212}=-k,\\
\label{eq27} ^FR_{0123} &=& ^FR_{0231}=-\frac{1}{2} q,\quad ^FR_{0312}=q;
\end{eqnarray} 
these are the only nonzero components except, of course, for the symmetries
of the Riemann tensor.
Here $(0,1,2,3)=(T,X,Y,Z)$, $k$ is given by equation \eqref{eq6} and $q$ is
directly proportional to the angular momentum of the source and is given by
\begin{equation}\label{eq28}
q=2\; GM a\frac{3r^2-a^2}{(r^2+a^2)^3}.
\end{equation}
One can interpret $q$ as the ``magnetic" curvature, while $k$ has the interpretation of
``electric" curvature. It is a significant feature of equations~\eqref{eq25}--\eqref{eq27}
that they are independent of the Lorentz factor of the reference trajectory $\gamma$.
This is a result of the fact that the axis of rotational symmetry in the Kerr metric 
provides two special tidal directions---corresponding to ingoing and 
outgoing trajectories---such that $k$ and $q$ are independent of any 
Lorentz boosts along these directions~\cite{9,10,11}. These special tidal
directions are related to the repeated principal null directions of
the curvature tensor and therefore the degenerate nature of Kerr spacetime.
That is, the special tidal directions exist at each event in Kerr
spacetime, 
since it is a vacuum spacetime of type $D$ in the 
Petrov classification~\cite{9,10,11}.
The significance of such directions in connection with the direction
of an astrophysical jet has been pointed out in~\cite{10}:
to preserve the collimation of a highly relativistic jet against
tidal disruptions caused by the central source, the most natural
jet direction would be a special tidal direction.
Moreover, of all the special tidal directions the Kerr rotation
axis has the additional advantage of being in conformity with the
symmetry of the configuration under consideration.

Using equations \eqref{eq25}--\eqref{eq27}, it is possible to
express equation \eqref{eq2} as 
\begin{eqnarray}
\label{eq29}
\nonumber &&\ddot{X}-\frac{1}{2}kX\big[1-2\dot{X}^2
+\frac{2}{3}(2\dot{Y}^2-\dot{Z}^2)\big] + \frac{1}{3} k\dot{X}( 5Y\dot{Y} -7Z\dot{Z})\\
&&\qquad + q[\dot{X}\dot{Y}\dot{Z} X-\dot{Z}Y(1+\dot{X}^2)-2\dot{Y}Z]=0,\\
\label{eq30}
\nonumber &&\ddot{Y}- \frac{1}{2} kY\big[ 1-2\dot{Y}^2 +\frac{2}{3}
(2 \dot{X}^2-\dot{Z}^2)\big] +\frac{1}{3} k\dot{Y} (5X\dot{X} -7Z\dot{Z})\\
&&\qquad-q [\dot{X}\dot{Y}\dot{Z}Y-\dot{Z}X(1+\dot{Y}^2)-2\dot{X}Z]=0,\\
\label{eq31}
\nonumber &&\ddot{Z} + kZ\big[ 1-2\dot{Z}^2+\frac{1}{3} (\dot{X}^2+\dot{Y}^2)\big] +\frac{2}{3}
k\dot{Z} (X\dot{X} +Y\dot{Y})\\
&&\qquad-q(X\dot{Y}-\dot{X} Y)(1-\dot{Z}^2)=0,
\end{eqnarray}
where $\dot{X}=dX/dT$, etc. 
These equations reduce to the one-dimensional generalized
Jacobi equation~\eqref{eq8} for $X=Y=0$.  
It is simple to check that equations~\eqref{eq29}--\eqref{eq31}
are invariant under the transformations $(X,Y,Z; k,q)\mapsto (Y,X,Z; k,-q)$
and $(X, Y, Z;k,q)\mapsto (X,Y,-Z;k,-q)$, since $q$ is directly related
to the rotation of the source.
Furthermore, equations~\eqref{eq29}--\eqref{eq31} are also invariant
under parity or time reversal if in each case we also change $q$ to $-q$.

It is interesting to note an important symmetry of the system~\eqref{eq29}--\eqref{eq31}
under rotations about the $Z$-axis. The rotational symmetry about the direction of
the motion of the reference particle implies that there is a degeneracy in the
choice of the $X$ and $Y$ axes of the tetrad frame of the reference observer. Once
an orthonormal tetrad is chosen, however, it is then parallel transported along
the reference worldline. 
Nevertheless, the axial symmetry of the configuration under
consideration implies that under a rotation about the $Z$-axis with a constant
azimuthal angle $\phi_0$,
\begin{eqnarray}
 \label{eq33a}       X' &=& X \cos \phi_0 + Y \sin \phi_0   ,\\            
  \label{eq34a}      Y' &=& - X \sin \phi_0 + Y \cos \phi_0                  
\end{eqnarray}
and $Z' = Z$,  the system~\eqref{eq29}--\eqref{eq31} remains invariant. 
This symmetry can, of course,
be checked directly; for instance, for $\phi_0 = \pi / 2$ we have $X' = Y$
 and $Y' = - X$ and
it is simple to see that the transformation $( X, Y ) \mapsto ( Y, - X )$ 
leaves the system~\eqref{eq29}--\eqref{eq31} invariant.

To express $k$ and $q$ as functions of the time $T$, we need
an explicit solution for the trajectory~\eqref{eq4}. To this end,
we set $\gamma=1$ as before and integrate
$dr/dT=\sqrt{2GMr}/\sqrt{r^2+a^2}$ with the initial condition
that $r=r_0$ at $T=0$, where $r_0^2> 3 a^2$.
We have thus far focused our attention on an ultrarelativistic
jet clump that is launched from $(X,Y,Z)=0$ at $T=0$ exactly along
the rotation axis of the source such that $X$ and $Y$ remain zero throughout
the flow.
What happens if the jet is \emph{not} launched \emph{exactly} along the $Z$-axis?
To answer this question, we must study the linearization of 
equations~\eqref{eq29}--\eqref{eq31} about a solution $Z=Z(T)$ of 
equation~\eqref{eq8}.

To obtain the linearization of  system~\eqref{eq29}--\eqref{eq31}
around the motion of an ultrarelativistic clump along the axis of symmetry,
we keep only terms linear in $X$, $\dot X$, $Y$ and $\dot Y$ 
and find that
\begin{eqnarray}
\label{eq33}
\ddot X-\frac{7}{3} k(Z\dot Z)\dot X-\frac{1}{2}k(1-\frac{2}{3} \dot Z^2)X
  &=& q(\dot Z Y+2 Z\dot Y),\\
\label{eq34}
\ddot Y-\frac{7}{3} k(Z\dot Z)\dot Y-\frac{1}{2}k(1-\frac{2}{3} \dot Z^2)Y
  &=& q(\dot Z X+2 Z\dot X),
\end{eqnarray}
while the third equation simply reduces to equation~\eqref{eq8}.

At any given time $T$, the new system~\eqref{eq33}--\eqref{eq34}
represents two damped harmonic oscillators that are coupled through the 
magnetic curvature $q$, i.e. the coupling is due to the rotation of the source;
moreover, the system decouples if it is transformed to the new variables
$X\pm Y$.
It is possible to obtain a simple analytic solution of this system that is
linearized about the solution~\eqref{eq16} of equation~\eqref{eq8}, which
amounts to $Z(T)\approx \mathcal{R}_0 w(1-w^2/6)$
with $w=T/ \mathcal{R}_0$. Writing
\begin{equation}\label{eq35}
k\approx -\frac{1}{\mathcal{R}_0^2}[1-3 w+O(w^2)],\quad
q\approx 3\frac{a}{r_0\mathcal{R}_0^2}[1-4 w+O(w^2)]
\end{equation}
for $0<w<1$, we find from equations~\eqref{eq33}--\eqref{eq34}
that 
\begin{eqnarray}
\label{eq36}
X(T) &=& \dot X(0) T F(\frac{T}{\mathcal{R}_0})+\frac{3}{2}\frac{a}{r_0}
\dot Y(0) T H(\frac{T}{\mathcal{R}_0}),\\
\label{eq37}
Y(T) &=& \dot Y(0) T F(\frac{T}{\mathcal{R}_0})+\frac{3}{2}\frac{a}{r_0}
\dot X(0) T H(\frac{T}{\mathcal{R}_0}),
\end{eqnarray}
where only terms linear in $a/r_0\ll 1$ are taken into account
and $F$ and $H$ are given by
\begin{equation}\label{eq38}
F(w)=1-\frac{5}{12}w^2+\frac{5}{8}w^3+O(w^4),\quad
H(w)=w^2-2w^3+O(w^4).
\end{equation}
These results, which are approximately valid for $\Gamma_0 \gg 1$ and
$T\ll\mathcal{R}_0$, provide insight into the general behavior of the
flow near the jet that moves along the $Z$-axis.
\begin{figure}
\vspace{1in}
\centerline{\psfig{file=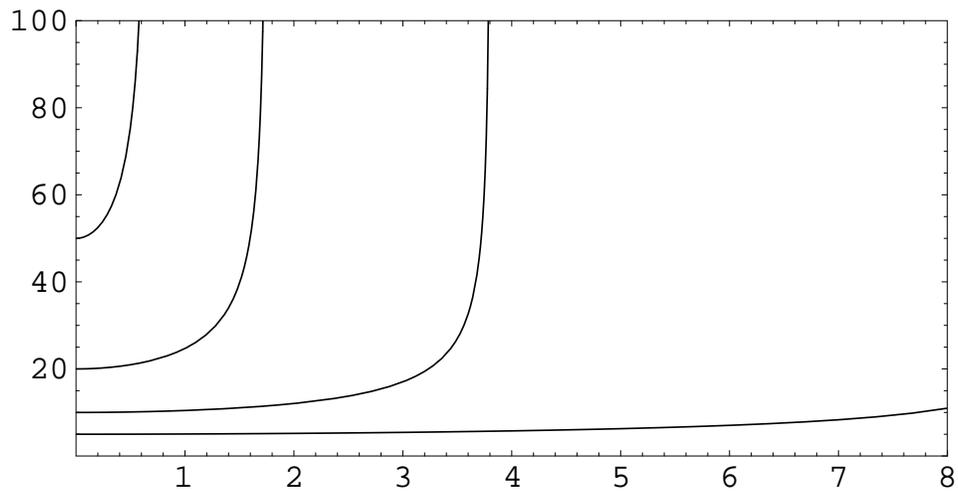, width=30pc}}
\caption{Plot of the Lorentz factor $\Gamma$ versus $T/(GM)$ based on
integration of equation~\eqref{mu43} with
initial data $X=0$ at $T=0$ with
$\dot X(0)=2\sqrt{6}/5$, $3\sqrt{11}/10$, $\sqrt{399}/20$ and 
$7\sqrt{51}/50$ corresponding respectively
to $\Gamma_0=5$, 10, 20 and 50. In this plot $a/(GM)=1$
and $r_0/(GM)=10$. The graph illustrates acceleration of
the particle such that $\Gamma$ essentially
approaches infinity at $T/(GM)\approx 3.9$, 1.8 and 0.7 
for $\Gamma_0=10$, 20 and 50, respectively.  \label{fig:7}}
\end{figure}
\begin{figure}
\vspace{1in}
\centerline{\psfig{file=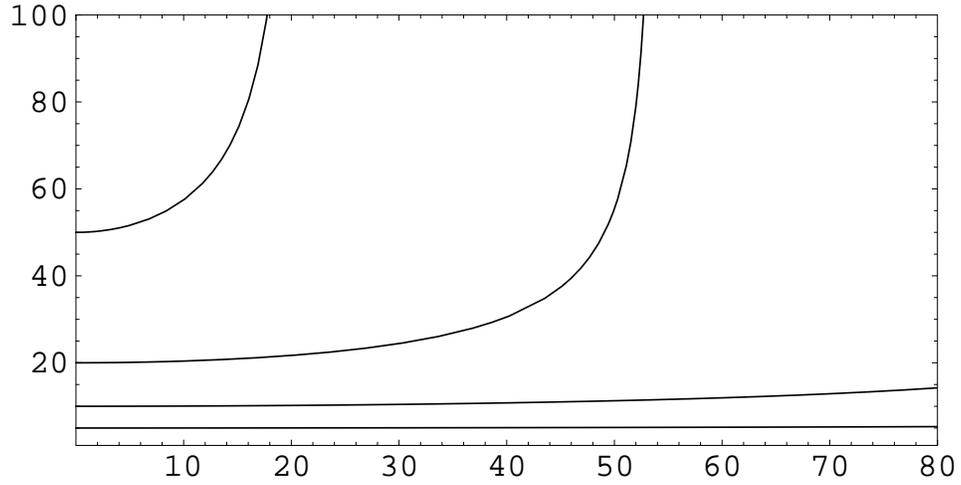, width=30pc}}
\caption{Plot of the Lorentz factor $\Gamma$ versus $T/(GM)$ based on
integration of equation~\eqref{mu43} with
initial data $X=0$ at $T=0$ with
$\dot X(0)=2\sqrt{6}/5$, $3\sqrt{11}/10$, $\sqrt{399}/20$ and 
$7\sqrt{51}/50$ corresponding respectively
to $\Gamma_0=5$, 10, 20 and 50. In this plot $a/(GM)=0$
and $r_0/(GM)=100$. The graph illustrates acceleration of
the particle such that $\Gamma$ essentially
approaches infinity at $T/(GM)\approx 55$ and 21 
for $\Gamma_0= 20$ and 50, respectively.  \label{fig:7b}}
\end{figure}

The axial symmetry of the system~\eqref{eq29}--\eqref{eq31} under rotation
about the $Z$-axis implies that it would be advantageous to express this
system in cylindrical coordinates.
This is done in~\ref{appen:A}, 
where an approximate constant of the motion 
for the system~\eqref{eq29}--\eqref{eq31}
is also derived based on axial symmetry. 

An important feature of equations~\eqref{eq29}--\eqref{eq31} is that
for $Y =Z = 0$, this system  reduces to 
\begin{equation}\label{mu43}
   \ddot X - \frac{1}{2} k X ( 1 - 2 \dot X^2 ) = 0.
\end{equation}      
This equation has the same form as equation~\eqref{eq8}
 except that the negative
curvature $k$ in~\eqref{eq8} is replaced by $ - k / 2$. 
It follows that a particle launched
from $( X, Y, Z ) = 0$ along the $X$-direction with a speed 
above the critical speed
$1/\sqrt{2}$ is \emph{accelerated} as in figures~\ref{fig:7} and~\ref{fig:7b}. 
It is a consequence of the axial symmetry
of the system~\eqref{eq29}--\eqref{eq31}  that the form of equation~\eqref{mu43}
is generic for motion along
any direction perpendicular to the $Z$-axis (see~\ref{appen:A}).

Equation~\eqref{mu43} again contains the critical speed $1/\sqrt{2}$.
In this approximation, 
if the initial speed is below this critical value, then the particle starting
with a positive initial velocity
decelerates over time intervals that are much longer than those
corresponding to the validity of the
approximation. While the asymptotic behavior of the system has no physical
meaning, one can show that the trajectories in the phase plane spiral
around the origin.

If the initial speed is above the critical value, then
the particle accelerates and its Lorentz factor approaches infinity in finite
time. In the latter case, it is interesting to extend our analytic approximation
scheme involving equations~\eqref{eq8}--\eqref{eq17} in section~\ref{s2}
for $\Gamma_0\gg 1$, $\Gamma\gg 1$ and $T\ll 2\mathcal{R}_0/3$ 
to equation~\eqref{mu43} by the simple substitution $k\mapsto -k/2$.
Instead of~\eqref{neweq17}, we get
\begin{equation}\label{mu:48}
\Gamma^{-2}\approx \Gamma_0^{-2}-\frac{1}{2}\Big(\frac{T}{\mathcal{R}_0}\Big)^2,
\end{equation}
which implies that $\Gamma$ diverges at 
\begin{equation}\label{mu:49}
\frac{T_d}{GM}\approx \frac{1}{\Gamma_0} \Big(\frac{r_0}{GM}\Big)^{3/2}.
\end{equation}
As expected, $T_d$ is inversely proportional to $\Gamma_0$ for
a fixed $r_0$.
For $\Gamma_0=10$, 20 and 50 in figure~\ref{fig:7}, equation~\eqref{mu:49}
predicts that $T_d/(GM)\approx 3.16$, 1.58 and 0.63, which compare favorably with
3.9, 1.8 and 0.7 obtained by the numerical integration used to produce
the figure. Similar results are obtained for  figure~\ref{fig:7b}; for instance,
$T_d/(GM)\approx 50$ and 20 for $\Gamma_0=20$ and 50, respectively,
which should be compared with 55 and 21 based on numerical results
presented in figure~\ref{fig:7b}.
The approximation~\eqref{mu:49} improves as $\Gamma_0$ increases.

The phase diagram for equation~\eqref{mu43} in the case of constant
curvature is given in figure 3 of~\cite{1}. It follows from this equation
that
\begin{equation}\label{mu52}
\frac{d}{dT}\ln\Big(\frac{1-2 V^2}{1-2 V_0^2}\Big)=-2k(T) V X,
\end{equation}
which can be obtained from~\eqref{eq11} by $k\mapsto -k/2$. The left
side of equation~\eqref{mu52} is positive; therefore, if
$V_0^2<1/2$, then $V^2$ decreases toward zero. 
Moreover, if $V_0^2>1/2$, then $V^2$ monotonically increases with time.

It is important to compare and contrast figures~\ref{fig:7} 
and~\ref{fig:7b}: the curvature is weaker for larger $r_0$; therefore,
for a given initial $\Gamma_0$, it takes longer for the
particle's speed to reach unity ($\Gamma\to\infty$).
In fact, $T_d^2\propto r^3_0$ from equation~\eqref{mu:49}, as
can be verified based on the data given for figures~\ref{fig:7} 
and~\ref{fig:7b}.

\section{Tidal acceleration}\label{s6}
\begin{figure}
\vspace{1in}
\centerline{\psfig{file=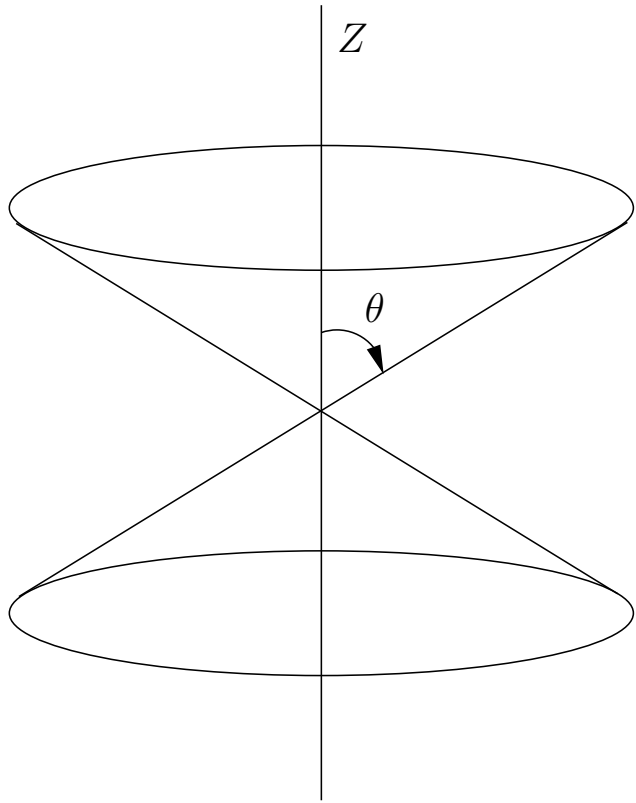, width=20pc}}
\caption{Schematic representation of the boundary cone separating 
the regions of relative deceleration and acceleration.
Particles whose initial velocities are inside the critical velocity
cone decelerate; particles
whose velocities are outside the cone accelerate.
The critical angle $\theta$ is such that $\cos^2 \theta =\frac{ 1}{3}$ and 
hence $P_2 (\cos \theta ) = 0$, where $P_2 (x) = ( 3 x^2 - 1 ) / 2$
 is the Legendre polynomial of the second order. 
This angle is known in physics as the ``magic angle" 
and occurs in several different contexts such as, for instance, 
NMR spectroscopy.
\label{fig:8}}
\end{figure}
We have demonstrated in the previous section that tidal acceleration
can occur in directions
perpendicular to the $Z$-axis if the speed of the clump exceeds the 
critical speed of $1/\sqrt{2}$. In the three-dimensional case, motion
relative to the standard reference trajectory can occur in $X$, $Y$
and $Z$ directions. The main factors responsible for reversing
the Newtonian behavior of the tides along these directions are
\begin{eqnarray}
\label{eq39}
\nu_X&=&\big[ 1-2\dot X^2+\frac{2}{3}(2 \dot Y^2- \dot Z^2)\big]_{T=0},\\
\label{eq40}
\nu_Y&=&\big[1-2\dot Y^2+\frac{2}{3}(2\dot X^2- \dot Z^2)\big]_{T=0},\\
\label{eq41}
\nu_Z&=&\big[1-2\dot Z^2+\frac{1}{3}(\dot X^2+ \dot Y^2)\big]_{T=0},
\end{eqnarray}
based on equations~\eqref{eq29}--\eqref{eq31}.
For instance, $\nu_X<0$ would imply that the character of the motion
along the $X$-direction could be opposite of that expected from the 
Newtonian gravitation theory.

Observe that $\nu_X$ and $\nu_Y$ cannot both be negative.
This follows from the inequality $\nu_X+\nu_Y>0$,
since
\begin{equation}\label{eq42}
\frac{1}{2}(\nu_X+\nu_Y)=\frac{1}{3}(1-\dot X^2-\dot Y^2)_{T=0}
+\frac{2}{3}(1-\dot Z^2)_{T=0},
\end{equation}
where the right side is a sum of two manifestly positive terms. 
Moreover, inspection of equations~\eqref{eq39}--\eqref{eq41}
reveals that 
\begin{equation}
\nu_X+\frac{5}{3}>0,\qquad \nu_Y+\frac{5}{3}>0,\qquad \nu_Z+1>0.
\end{equation}
On the other hand, it is in general possible to have $\nu_Z<0$ and
either $\nu_X$ or $\nu_Y$ also negative. An example of this
situation is provided by
the motion of a particle starting from the origin at $T = 0$
with velocity components
given by $\dot X=\sqrt{19/60}$, $\dot Y=0$
and $\dot Z=\sqrt{5}/3$. While the dynamics of this case is complicated,
numerical experiments show that $\Gamma$ slowly decreases with time.

It follows from these results that highly relativistic particles
emitted near a gravitationally collapsed configuration are accelerated along 
directions normal to the rotation axis and hence immediately leave the black hole environment,
while highly relativistic particles emitted along the rotation axis
are decelerated and appear as jets. If the collapsed configuration is 
nonrotating, then deceleration occurs along the radial direction and 
acceleration occurs in directions normal to the radial direction.

We have performed a series of numerical experiments to delineate
the regions of acceleration and deceleration of particles relative
to the reference particle. The results are represented in figure~\ref{fig:8}.
For $r_0/(GM)=100$, numerical experiments show that $\theta$ is independent
of $\Gamma_0$ and is
given approximately by $\tan\theta=\sqrt{2}$, so that $\theta\approx 54.7^\circ$.
In our numerical experiments, $\theta$ remained approximately
constant for $\Gamma_0\ge 5$ and 
$5\le r_0/(GM)\le 100$.

To see how this comes about from an approximate analytical point of view, we
consider equations~\eqref{eq29}--\eqref{eq31} in cylindrical coordinates, 
namely, equations~\eqref{mu:44}--\eqref{mu:46}. It follows
from our numerical experiments that the influence of the rotation 
of the source on
the relative motion is generally negligible.  
To simplify matters, 
the rotation of the source can therefore be neglected and we then
confine our considerations to the $(\rho,Z)$-halfplane, 
since $\dot\varphi = 0$ is a
solution of equation~\eqref{mu:45} with $q = 0$. 
Thus the equations of motion reduce to
\begin{eqnarray}
\label{bn:53}
\ddot\rho - \frac{1}{2} k \rho&=&-k Q_\rho ,\\
\label{bnx:53} \ddot Z + k Z&=&- k Q_Z,
\end{eqnarray}   
where $-k > 0$ and 
\begin{eqnarray}
\label{bn:54}
 Q_\rho &=& \rho ( \dot\rho^2 +\frac{1}{3} \dot Z^2 ) 
   - \frac{7}{3} \dot\rho Z \dot Z,\\
\label{bnx:54}
Q_Z &=& Z ( \frac{1}{3} \dot\rho^2 - 2 \dot Z^2 ) + \frac{2}{3}
    \rho \dot\rho\dot Z.             
\end{eqnarray}
If we set $Q_\rho = Q_Z = 0$, equations~\eqref{bn:53} and~\eqref{bnx:53} 
reduce to the standard Jacobi equation;
therefore, the special deceleration/acceleration phenomena 
discussed in this work
come about as a consequence of the existence of nonzero $Q_\rho$ and $Q_Z$. 

Let us now consider a particle starting out at time $T = 0$  
from the position of the reference particle at $\rho= Z = 0$. We 
will predict the motion of the test particle,
determined by equations~\eqref{bn:53} and~\eqref{bnx:53}, in case its 
initial velocity vector is 
$(\dot \rho, \dot Z ) = (V_0 \sin \theta, V_0 \cos \theta )$, where
$0 < V_0 < 1$ and the initial velocity vector makes an angle $\theta$ 
with respect to the $Z$-axis.
To do this, we let the function $\alpha$ be defined by 
\begin{equation}\label{bn:57}
\rho(T) = \alpha ( T ) Z(T),
\end{equation}           
substitute equation~\eqref{bn:57} 
in equation~\eqref{bn:53} and use equation~\eqref{bnx:53} to obtain
the differential equation
\begin{equation}\label{bn:58}
\ddot \alpha Z + 2 \dot\alpha\dot Z 
 - \frac{3}{2} \alpha k Z 
+ \frac{1}{3}\dot\alpha k Z^2 ( 2 \alpha \dot\rho - 7 \dot Z ) = 0 . 
\end{equation} 
The Taylor expansions
\begin{eqnarray}
\label{bn:59}
\alpha ( T ) &=& ( 1 + \alpha_1 T + \frac{1}{2} \alpha_2 T^2 +\cdots ) \tan \theta ,\\           
\label{bn:60} Z(T) &=& (V_0 \cos \theta) T +\cdots,\\                                                   
\label{bn:61} k(T)&=& - \mathcal{R}_0^{-2} 
( 1 - \frac{3}{\mathcal{R}_0} T+ \cdots ),                           
\end{eqnarray}
may be employed in equation~\eqref{bn:58} to show that 
\begin{equation}\label{bn:62}
\alpha_1 = 0 ,\qquad \alpha_2 = -\frac{1}{2}\mathcal{R}_0^{-2}    . 
\end{equation}
Thus for $T\ll\mathcal{R}_0$, 
neglecting terms of second order and higher in $T/\mathcal{R}_0$
compared to unity, the particle trajectory is such that
\begin{equation}\label{bn:63}
  \rho = Z \tan \theta,\qquad    \dot \rho = \dot Z \tan \theta   
\end{equation} 
are approximately valid. Plugging these relations in 
equations~\eqref{bn:54} and~\eqref{bnx:54}, we
find that for $T\ll \mathcal{R}_0$,
\begin{eqnarray}
\label{bn:64}
Q_\rho &=& \rho \dot Z^2 ( \tan^2 \theta - 2 ),\\ 
\label{bn:65}
Q_Z &=& Z \dot Z^2 ( \tan^2 \theta - 2 ).   
\end{eqnarray}
For $\tan^2 \theta < 2$, we have deceleration, 
while for $\tan^2 \theta > 2$, we have
acceleration according to equations~\eqref{bn:53} and~\eqref{bnx:53}. 
Thus the boundary region is given by $\tan^2 \theta =
2$, as illustrated in figure~\ref{fig:8}.

\section{Discussion}\label{s7}
In this paper, we are interested in some of the observable consequences of
general relativity involving ultrarelativistic flows near a gravitationally
collapsed configuration. The measurement problem in general relativity is
rather subtle. Suppose, for instance, that in the background of Kerr spacetime
one succeeds in solving certain covariant equations of 
magnetohydrodynamics (MHD) in Boyer-Lindquist coordinates. 
Though possibly interesting, this result
per se may have little to do with understanding the \emph{observable} 
consequences of
general relativistic MHD in this case. As a first step, we therefore
concentrate on an \emph{invariant} characterization of the dynamics of
free particles
in the exterior Kerr spacetime; in particular, we study \emph{relative}
motion in the
simplest possible situation, namely, the motion of free particles relative to
a certain fiducial class of particles moving freely on escape trajectories
along the rotation axis of the Kerr source. The  relative motion of
two nearby free particles is then determined by the generalized Jacobi equation.

The generalized Jacobi equation has been known for three decades~\cite{5},
but its physical consequences for relativistic relative motion  have
only recently received attention~\cite{1,2}.
An important issue regarding the generalized Jacobi equation is that for 
ultrarelativistic relative motion, the duration of validity of this 
equation is rather short. However, we have shown that even
within this limit, for ultrarelativistic outward motion along
the rotation axis of a Kerr black hole, there is a
remarkably strong initial tidal deceleration relative to the ambient medium regardless
of the value of $\Gamma _0\gg 1$. To go beyond this initial deceleration, let us note
that a significant feature  of our basic equations~\eqref{eq4}--\eqref{eq6} 
is that they do not  explicitly depend on the reference
trajectory corresponding to a  ``fixed" marker in the ambient medium. One can therefore
imagine a  number of such markers along the path of the jet such that from one 
marker to the next, the integration of equation \eqref{eq8} describes 
the force-free jet deceleration to a good approximation. In this way,  one may 
approach the terminal speed of the jet by referring the jet motion to 
different markers. So long as the jet speed relative to any nearby 
ambient marker is significantly greater than $1/\sqrt{2}$, it tends to decrease 
toward this terminal speed.

It is possible to view the terminal speed in another way. Along a radial direction far
from a massive source, its tidal influence on two neighboring test particles is a tidal
{\it acceleration} if the relative speed of the particles is significantly below
$1/\sqrt{2}$. However, this influence turns to a tidal {\it deceleration} if the
relative speed is significantly above $1/\sqrt{2}$. 

We have neglected plasma effects in this 
paper; in fact, our treatment is valid for a force-free plasma. If the 
plasma is not force-free, the inclusion of its effects would 
enormously complicate matters \cite{2,8}. In any case, tidal 
deceleration is most effective very close to the central source, where the 
jet originates. Moreover, it is expected to be more significant for 
microquasars, since tidal effects near a black hole
 generally decrease as the inverse 
square of the mass of the source. Thus the results of this paper may 
well play a significant role in the dynamics of jets, 
especially in microquasars~\cite{4}. More generally, this work may be of interest in connection with the
astrophysics of energetic particles such as the ultrahigh energy cosmic rays.

Let us now turn to the motion of a particle normal to the rotation axis of the
black hole. Relative to the reference particle, the particle accelerates if its
initial speed exceeds the critical speed $1/\sqrt{2}$. For ultrarelativistic
relative motion, this tidal acceleration can lead to ultrahigh 
energy particles according to the generalized Jacobi equation; more
generally, particle acceleration occurs outside a cone of angle $\theta$, where
the polar angle $\theta$ is measured from the $Z$-axis and 
$\tan \theta \approx \sqrt{2}$.
As the particle accelerates beyond the Fermi system, 
one can in principle use other
overlapping coordinate systems to describe the motion 
over an extended period of
time. On the other hand, the tidal effects of the black hole 
diminish rapidly with
increasing distance and hence the influence of the black hole 
can be neglected once
the particle has moved sufficiently far away. 

The rotation of the source leads to a preferred radial direction, i.e.
the axis of rotation, but is otherwise of little importance for the
tidal dynamics presented in this paper. This remark applies equally well 
to the exact nature of the source, except that our
results become physically significant near gravitationally 
collapsed objects, where
tidal accelerations can be substantial.

The nature of the accretion mechanisms that create ultrarelativistic particles
near black holes is beyond the scope of our investigation. The analysis
presented in this paper based on the generalized Jacobi equation indicates that
once such particles are created near the poles of the black hole, those propagating mainly along the rotation
axis of the black hole decelerate with respect to the ambient medium, while
those propagating mainly normal to this axis can accelerate to
almost the speed of light. 

The loss of kinetic energy for the motion of an initially ultrarelativistic
particle along the rotation axis and the gain in kinetic energy for motion
normal to this axis may be attributed, in analogy with Newtonian gravitation, to
the change in the gravitational potential energy, which is the tidal energy in
the case under consideration in this paper. The underlying assumption here is
that the perturbation of the background geometry caused by the motion of the
test particle may be neglected. This breaks down, however, 
when the particle is accelerated to almost the speed of light; 
therefore, the back-reaction on the motion is expected to moderate
this singularity leading to an ultrahigh energy particle.
More generally, the gravitational and electromagnetic 
(in case of charged particles) radiations emitted by 
the decelerating/accelerating particles should be taken into account as well.
The tidal acceleration mechanism may provide the key to the explanation of 
recent \emph{Chandra} X-ray observations of the Crab Nebula.

It is well known that primary ultrahigh energy cosmic 
ray protons with energies above
$10^{20}$ eV from extragalactic sources are expected to interact with the cosmic
microwave background photons resulting in photopion 
production and pair creation~\cite{13,14}.
Therefore, it may be reasonable to assume that the most energetic
particles that reach the solar system originate within our galaxy 
(see~\cite{15} for
a recent review of cosmic ray astrophysics). In view of the foregoing results, 
it would be interesting to determine whether the directional distribution of the
ultrahigh energy cosmic rays is correlated with the known microquasars, taking
due account of the directionality of the acceleration zones normal to the jet
directions.

\appendix
\newcounter{saveeqn}%
\newcommand{\alpheqn}{\setcounter{saveeqn}{\value{equation}}%
\stepcounter{saveeqn}\setcounter{equation}{0}%
\renewcommand{\theequation}
  {\mbox{\Alph{section}\arabic{equation}}}}%
\alpheqn
\renewcommand{\thesection}{Appendix \Alph{section}}
\section{Cylindrical coordinates}\label{appen:A}
In system~\eqref{eq29}--\eqref{eq31}, let us introduce the 
cylindrical coordinates
$\rho$, $\varphi$ and $Z$, where
\begin{equation}\label{mu:43}
X=\rho\cos\varphi, \qquad Y=\rho\sin\varphi.
\end{equation}
The resulting system is given by 
\begin{eqnarray}
\label{mu:44}
\nonumber &&
\ddot\rho-\rho\dot\varphi^2-\frac{1}{2}k[\rho(1-2\dot\rho^2
+\frac{4}{3}\rho^2\dot\varphi^2-\frac{2}{3}\dot Z^2)
+\frac{14}{3}\dot\rho Z\dot Z]\\
&&\qquad +q\rho \dot\varphi(\rho\dot\rho\dot Z-2Z)=0,\\
\label{mu:45} 
&&
\rho\ddot\varphi+2\dot\rho\dot\varphi
+\frac{1}{3}k\rho\dot\varphi(5\rho\dot\rho-7 Z\dot Z)
+q[2\dot\rho Z+\rho\dot Z(1+\rho^2\dot\varphi^2)]=0,\quad\\
\label{mu:46} 
&&
\nonumber \ddot Z+k\{Z[1+\frac{1}{3}(\dot\rho^2+\rho^2\dot\varphi^2)-2\dot Z^2]
+\frac{2}{3}\rho\dot\rho\dot Z\}\\
&&\qquad-q\rho^2\dot\varphi(1-\dot Z^2)=0.
\end{eqnarray}
This system depends on $\varphi$ only through $\dot\varphi$ and
$\ddot\varphi$ and is therefore
 invariant under $\varphi\mapsto \varphi+\text{constant}$, 
which again reflects the axial symmetry of the system. 

A remarkable feature of system~\eqref{mu:44}--\eqref{mu:46} is that
for $\dot \varphi=0$ and $Z=0$, the system reduces to purely radial
motion given by 
\begin{equation}\label{mu:47}
\ddot\rho-\frac{1}{2}k\rho(1-2\dot\rho^2)=0,
\end{equation} 
which is the same form as equation~\eqref{eq8} except for
$k\mapsto -k/2$. This reversal of sign of the curvature leads to the
\emph{acceleration} of an ultrarelativistic particle that is launched
from $\rho=0$ in a direction normal to the $Z$-axis as illustrated in
figures~\ref{fig:7} and~\ref{fig:7b}. 

It has been shown in~\cite{1} that the tidal equation is derived 
from a Lagrangian given in Fermi coordinates by
\begin{equation}\label{a6}
L=-\sqrt{-g_{00}-2 g_{0i}V^i-g_{ij}V^iV^j}\, ,
\end{equation}
where the metric components in Fermi coordinates are given by
equation~\eqref{metrict}. Using equations~\eqref{eq25}--\eqref{eq27},
it is possible to show that in our approximation the Lagrangian reduces to
\begin{equation}\label{a7}
\tilde L=-\sqrt
{1-\dot X^2-\dot Y^2-\dot Z^2-\frac{1}{2}k(T) S-2q(T)(X\dot Y-\dot XY)Z}\, ,
\end{equation}
where $S(X,Y,Z,\dot X,\dot Y, \dot Z)$ is given by
\begin{equation}\label{a8}
S=X^2+Y^2-2 Z^2+\frac{2}{3}(X\dot Y-\dot X Y)^2
-\frac{1}{3}(X\dot Z-\dot X Z)^2-\frac{1}{3}(Y\dot Z-\dot Y Z)^2.
\end{equation}
Expressed in cylindrical coordinates, $\tilde L$ takes the form
\begin{equation}\label{a9}
\tilde L=-\sqrt
{1-\dot\rho^2-\rho^2 \dot\varphi^2-\dot Z^2
-\frac{1}{2}k(T) \Psi-2q(T)\rho^2\dot\varphi Z}\, ,
\end{equation}
where 
\begin{equation}\label{a10}
\Psi=\rho^2-2 Z^2+ \frac{1}{3}\rho^2(2 \rho^2-Z^2)  \dot\varphi^2
-\frac{1}{3} (\rho\dot Z-\dot\rho Z)^2.
\end{equation}
Here $\varphi$ is a cyclic coordinate; therefore,
\begin{equation}\label{a11}
P_\varphi=\frac{\partial \tilde L}{\partial\dot\varphi}
\end{equation}
is an approximate constant of the motion for system~\eqref{mu:44}--\eqref{mu:46}
given by
\begin{equation}\label{a12}
P_\varphi=\frac{[1+\frac{1}{6}k(T)(2 \rho^2-Z^2)]\rho^2\dot\varphi+q(T)\rho^2 Z}
 {\sqrt
{1-\dot\rho^2-\rho^2 \dot\varphi^2-\dot Z^2
-\frac{1}{2}k(T) \Psi-2q(T)\rho^2\dot\varphi Z}}\,.
\end{equation}
One can interpret $P_\varphi$ as a generalized orbital angular
momentum about the $Z$-axis; its approximate constancy in time is related to 
the axial symmetry of the system~\eqref{mu:44}--\eqref{mu:46}.
We do not know if system~\eqref{mu:44}--\eqref{mu:46} is Lagrangian.

\section{Higher-order tidal accelerations}\label{appen:B}
\setcounter{equation}{0}
\begin{figure}
\centerline{\psfig{file=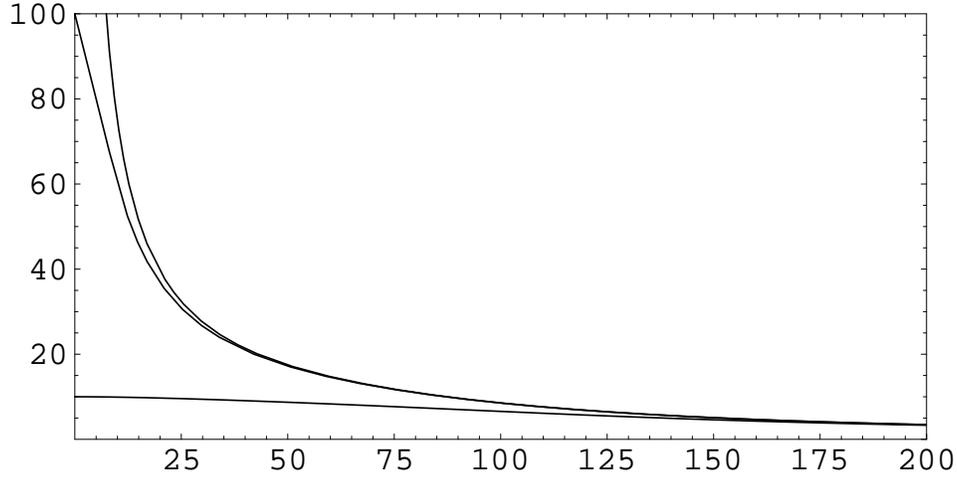, width=30pc}}
\caption{Plot of the Lorentz factor $\Gamma$ versus $T/(GM)$  
based on the integration
of equation~\eqref{B1} with initial data
$r_0/(GM)=100$, $Z(0)=0$ and $V(0)=\sqrt{\Gamma_0^2-1}\,/\,\Gamma_0$ for 
$a/(GM)=0$. The graph illustrates the deceleration of clumps with
initial Lorentz factors $\Gamma_0=1000$, 100 and 10. 
\label{newnewfig:6}}
\end{figure}
\begin{figure}
\centerline{\psfig{file=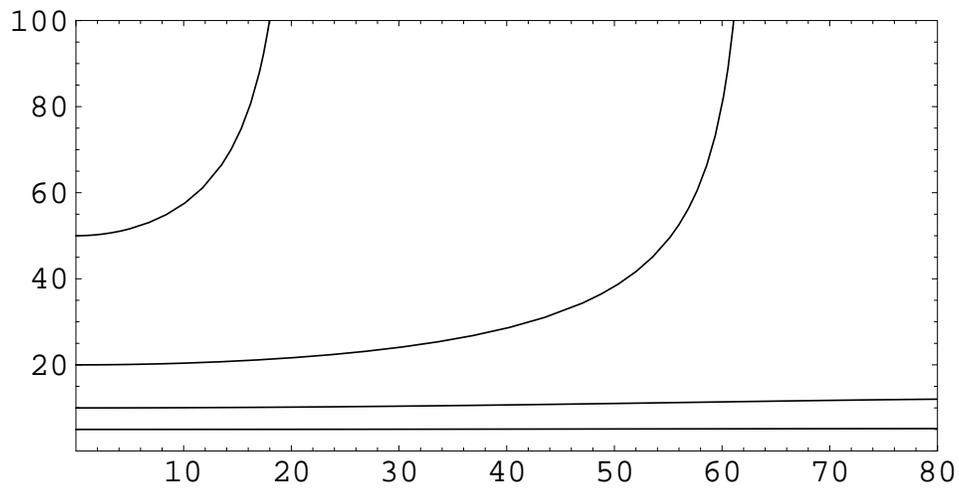, width=30pc}}
\caption{Plot of the Lorentz factor $\Gamma$ versus $T/(GM)$ based on
integration of equation~\eqref{B13} with
initial data $X=0$ at $T=0$ with
$\dot X(0)=2\sqrt{6}/5$, $3\sqrt{11}/10$, $\sqrt{399}/20$ and 
$7\sqrt{51}/50$ corresponding respectively
to $\Gamma_0=5$, 10, 20 and 50. In this plot $a/(GM)=0$
and $r_0/(GM)=100$. The graph illustrates acceleration of
the particle such that $\Gamma$ essentially
approaches infinity at $T/(GM)\approx 63$ and 21 
for $\Gamma_0=20$ and 50, respectively. 
\label{newnewfig:7}}
\end{figure}

It is in principle possible to obtain the 
infinite set of higher-order tidal accelerations 
that have been ignored in our analysis thus far.
Moreover, we must impose the requirement that for 
a timelike geodesic $g_{00} + 2 g_{0i} V^i + g_{jk} V^j V^k < 0$.
The generalized Jacobi equation contains the first-order
tidal terms~\cite{5}. 
Tidal terms of second and third order were first calculated in~\cite{6}
and~\cite{7}, respectively. Based on these
previous detailed calculations, the first two such 
contributions to equation \eqref{eq5} can be computed and hence the 
resulting augmented form of equation \eqref{eq5} is
\begin{eqnarray}\label{B1}
\nonumber \frac{d^2Z}{dT^2} &=&-(k+\frac{1}{2}k'Z+\frac{1}{6}k''Z^2)
(1-2 V^2)Z+ \frac{1}{2}(K+\frac{1}{3}K' Z) VZ^2\\
&&{}- \frac{2}{3}k^2(1+V^2)Z^3+O(Z^4).
\end{eqnarray}
Here $V=dZ/dT$,
$k$ is given by equation \eqref{eq6} as before and
\begin{eqnarray}
\label{B2}
k'(T)&=&\fr_{0303\,;\,3}=R_{\mu\nu\rho\sigma\,;\,\delta}\,
\lambda_{(0)}^\mu \lambda_{(3)}^\nu \lambda_{(0)}^\rho
  \lambda_{(3)}^\sigma\lambda_{(3)}^\delta,\\
\label{B3}
k''(T)&=&\fr_{0303\,;\,33}=R_{\mu\nu\rho\sigma\,;\,\delta\omega}\,
\lambda_{(0)}^\mu \lambda_{(3)}^\nu \lambda_{(0)}^\rho
  \lambda_{(3)}^\sigma\lambda_{(3)}^\delta\lambda_{(3)}^\omega,\\
\label{B4}
K(T)&=&\fr_{0303\,;\,0}=R_{\mu\nu\rho\sigma\,;\,\delta}\,
\lambda_{(0)}^\mu \lambda_{(3)}^\nu \lambda_{(0)}^\rho
  \lambda_{(3)}^\sigma\lambda_{(0)}^\delta,\\
\label{B5}
K'(T)&=&\fr_{0303\,;\,30}=R_{\mu\nu\rho\sigma\,;\,\delta\omega}\,
\lambda_{(0)}^\mu \lambda_{(3)}^\nu \lambda_{(0)}^\rho
  \lambda_{(3)}^\sigma\lambda_{(3)}^\delta\lambda_{(0)}^\omega,
\end{eqnarray}
along the reference worldline where 
$\partial x^\mu/\partial X^\alpha=\lambda^\mu_{\;\;(\alpha)}$.
In principle, the projections of the
covariant derivatives of the Riemann curvature tensor of Kerr spacetime
on the tetrad frame of the reference particle evaluated
along its worldline are required for the determination of the 
higher-order tidal accelerations.
However, for $r_0\gg GM$,
it is simpler and more instructive to employ the linear approximation
of general relativity as follows. The linearized Kerr metric
in the post-Newtonian framework is given by
\begin{eqnarray}\label{B6}
\nonumber ds^2&=&-(1+2\Phi)\,dt^2+\frac{4GJ}{r^3}(y\,dx-x\,dy)\,dt\\
&&{}+(1-2\Phi)(dx^2+dy^2+dz^2),
\end{eqnarray}
where $\Phi=-GM/r$ is the Newtonian potential. 
This metric can be written as $g_{\mu\nu}=\eta_{\mu\nu}+h_{\mu\nu}$,
where $h_{00}=-2 \Phi$, $h_{ij}=-2\Phi\delta_{ij}$ and
\begin{equation}\label{B7}
h_{0i}=-2G\frac{(\mathbf{J}\times\mathbf{r})_i}{r^3}
\end{equation}
is proportional to the gravitomagnetic vector potential. 
The Riemann tensor in the linear approximation is given by
\begin{equation}\label{B8}
R_{\mu\nu\rho\sigma}=\frac{1}{2}
(h_{\mu\sigma\,,\,\nu\rho}+h_{\nu\rho\,,\,\mu\sigma}
-h_{\nu\sigma\,,\,\mu\rho}-h_{\mu\rho\,,\,\nu\sigma}),
\end{equation}
which can be represented in the standard manner as a symmetric 
$6\times 6$ matrix with indices that range over the set
$\{01, 02, 03, 23, 31, 12\}$. For the linearized Kerr metric~\eqref{B6},
we have such a representation in terms of symmetric and traceless 
$3\times 3$ matrices $E$ and $B$,
\begin{equation}\label{B9}
\left[ \begin{array}{cc}
E & B \\
 B & -E
\end{array}\right ],
\end{equation}
where $E$ and $B$ are the electric and magnetic parts of the curvature
and 
\begin{eqnarray}
\label{B10}
E_{ij}&=& R_{0i0j}=\Phi_{,\,ij}
=\frac{GM}{r^3}(\delta_{ij}-3\frac{x^ix^j}{r^2}),\\
\label{B11}
B_{ij}&=& \frac{1}{2}\epsilon_{jk\ell}R_{0ik\ell}=
-3\frac{G}{r^5}[x^iJ^j+x^jJ^i+(\delta_{ij}-5\frac{x^ix^j}{r^2})
\mathbf{r}\cdot\mathbf{J}].
\end{eqnarray}

In the \emph{linear} approximation under consideration here
$\lambda^\mu_{\;\:(\alpha)}$ along the reference worldline in effect
reduces to  $\delta^\mu_{\;\:\alpha}$ in the computation of the
Riemann tensor and its covariant derivatives 
(which reduce to partial derivatives)
as measured by the reference particle. 
It is therefore interesting to note that along the reference
trajectory $(x=y=0,\, z=r)$,
the Riemann tensor measured by the reference particle turns
out to be identical to equations~\eqref{eq25}--\eqref{eq27}
without the $a^2$ terms,
as can be verified by direct calculation using~\eqref{B10} and~\eqref{B11}.

The rotation of the source plays a rather minor role in
the numerical results of this paper;
therefore, for the sake of simplicity we will set $J=0$ in
the computation of 
$k'$, $k''$, $K$ and $K'$.
We find that $K=K'=0$
and $k=-2GM/r^3$,
\begin{equation}\label{B12}
k'=6\frac{GM}{r^4},\qquad k''=-24\frac{GM}{r^5},
\end{equation}
so that $k'=k_{\,,\,3}$ and $k''=k_{\,,\,33}$ along the reference
path $(z=r)$.
It follows that the results in equation~\eqref{B12} agree with the
higher-order Newtonian tidal terms in equation~\eqref{eq19a}.
Thus the first term on the right-hand side of~\eqref{B1} contains
the higher-order Newtonian tidal terms once $1-2V^2\mapsto 1$ in the
nonrelativistic limit.
Equation~\eqref{B1} can now be integrated with $r=r(T)$ given
by~\eqref{eq9}; the results of this integration are presented
in figure~\ref{newnewfig:6}, where the Lorentz factor versus $T$ is given 
for $r_0/(GM)=100$ and initial Lorentz factors 1000, 100 and 10.
This figure should be compared and contrasted with figure~\ref{fig:1a}.

Let us now turn to the higher-order tidal corrections to equation~\eqref{mu43}.
Based on the detailed treatments in~\cite{6,7}, the result is
\begin{eqnarray}
\label{B13}
\nonumber \frac{d^2X}{dT^2}&=&-X(1-2\dot X^2)(\fr_{0101}+\frac{1}{2!}\fr_{0101\,;\,1}X
        +\frac{1}{3!}\fr_{0101\,;\,11}X^2)\\
\nonumber &&{}+\frac{1}{2}X^2\dot X(\fr_{0101\,;\,0}+\frac{1}{3}\fr_{0101\,;\,10}X)
-\frac{2}{3}X^3(1+\dot X^2)(\fr_{0101})^2\\
&&{}+O(X^4).
\end{eqnarray}
Using our approximation scheme with $J=0$,
we find that along the trajectory $\fr_{0101\,;\,0}=\fr_{0101\,;\,10}=0$
and 
\begin{eqnarray}\label{B14}
\fr_{0101}=\frac{GM}{r^3},\quad \fr_{0101\,;\,1}=0,\quad
\fr_{0101\,;\,11}=-9\frac{GM}{r^5}.
\end{eqnarray}
Equation~\eqref{B13} can be integrated with $r=r(T)$ given by~\eqref{eq9};
the Lorentz factor versus $T$ is given in figure~\ref{newnewfig:7} for
$r_0/(GM)=100$. 
The results should be compared and contrasted with those 
given in figure~\ref{fig:7b}.
The terms in equations~\eqref{B1} and~\eqref{B13} proportional
to $k^2$ make negligible contributions to the numerical work presented
in figures~\ref{newnewfig:6} and~\ref{newnewfig:7}, respectively,
in agreement with our approximation scheme.
 
Within the limit of validity of the tidal
equation, 
the higher-order tidal terms provide small corrections
to the generalized Jacobi equation as in
figures~\ref{newnewfig:6} and~\ref{newnewfig:7}. 
To demonstrate that the main results of this paper, namely, the strong initial
tidal deceleration of ultrarelativistic flows 
along the rotation axis and the acceleration normal to this axis 
will remain unchanged by the addition of
the higher-order tidal terms, we have integrated equations~\eqref{B1}
and~\eqref{B13} with the same
initial data as in figures~\ref{fig:1a} and~\ref{fig:7b}, respectively. 
The results are presented in
figures~\ref{newnewfig:6} and \ref{newnewfig:7}, respectively, 
which appear to be almost
identical to figures~\ref{fig:1a} and \ref{fig:7b}, respectively
and demonstrate that the higher-order tidal terms slightly enhance
the deceleration effect and tend to moderate the acceleration effect. 
For instance, in figure~\ref{fig:1a}, the graph for
initial 
$\Gamma_0=100$ reaches $\Gamma \approx 4.63$ at $T=100\; GM$, while in 
figure~\ref{newnewfig:6} it
reaches $\Gamma \approx 3.44$. Nevertheless, figures~\ref{newnewfig:6} 
and~\ref{newnewfig:7} demonstrate
that the inclusion of higher-order tidal terms will not change the main results
of this work.

Concerning the tails in figures~\ref{fig:1}, \ref{fig:1a} and \ref{newnewfig:6}, we note that so long as
the initial $\Gamma_0$ is well above the critical value of $\sqrt{2}$, a temporal
interval of integration can be found such that the flow decelerates as described in
detail in this paper. However, the size of this temporal interval of validity decreases
as the flow speed approaches the critical speed. Thus sufficiently close to the
critical speed our approach loses its validity. Moreover, based on the available terms
of the tidal equation
\eqref{B1}, it is clear that the critical speed $1/\sqrt{2}$ is {\it not} a feature
of the general tidal equation in this case. It follows that our treatment breaks down
when the jet clump decelerates to a speed sufficiently close to this critical speed. On
the astrophysical side, we note that by the time the clump decelerates close to the
critical speed relative to the ambient medium, the black hole may be sufficiently far
from the clump that tidal effects may no longer play a significant role in the
subsequent motion of the clump. 

\newcommand{\reseteqn}{\setcounter{equation}{\value{saveeqn}}%
\renewcommand{\theequation}{\arabic{equation}}}%

\end{document}